\documentclass[preprint,12pt]{elsarticle}
\pdfoutput=1
\usepackage[utf8]{inputenc}
\usepackage{a4wide}
\usepackage{amsfonts}
\usepackage{amssymb}
\usepackage{amsmath}
\usepackage{graphicx}
\usepackage{booktabs}
\usepackage{array}
\usepackage{color}
\usepackage{ulem}
\usepackage[small,bf]{caption}
\usepackage[usenames,dvipsnames]{xcolor}
\usepackage{subfigure}
\usepackage{verbatim}
\usepackage{dsfont}
\usepackage{bbm}
\usepackage{hyperref}
\usepackage{listings}
\usepackage{titlesec}
\usepackage{cleveref}
\usepackage{xspace}
\usepackage{textcomp}

\newcommand{\tapir}{\texttt{tapir}\xspace}
\newcommand{\qgraf}{\texttt{qgraf}\xspace}
\newcommand{\FORM}{\texttt{FORM}\xspace}
\newcommand{\python}{\texttt{python}\xspace}
\newcommand{\UFO}{\texttt{UFO}\xspace}
\newcommand{\UFOlong}{\texttt{Universal FeynRules Output}\xspace}
\newcommand{\UFOReader}{\texttt{UFOReader}\xspace}

\definecolor{codegreen}{rgb}{0,0.6,0}
\definecolor{codepurple}{rgb}{0.58,0,0.82}
\definecolor{backcolour}{rgb}{0.85,0.95,1.0}

\lstdefinestyle{conffilestyle}{
    backgroundcolor=\color{backcolour},   
    commentstyle=\color{codegreen},
    keywordstyle=\color{magenta},
    stringstyle=\color{codepurple},
    basicstyle=\ttfamily,
    breakatwhitespace=false,         
    breaklines=true,                 
    captionpos=b,                    
    keepspaces=true,                                   
    showspaces=false,                
    showstringspaces=false,
    showtabs=false,                  
    tabsize=2,
}

\lstdefinelanguage{shell}
{
  morekeywords={
    \$
  },
  sensitive=false, 
  morecomment=[l][\color{codegreen}]{-},
  morestring=[b]",
}

\lstdefinelanguage{form}
{
  morekeywords={
    id,
    if,
    else,
    elsif,
    endif,
    occurs,
    exit,
    repeat,
    endrepeat,
  },
    otherkeywords={
     \#define
    },
  sensitive=false, 
  morecomment=[f][\color{gray}][0]{*},
  morecomment=[f][\color{magenta}][0]{*--\#[},
  morecomment=[f][\color{magenta}][0]{*--\#]},
  morecomment=[f][\color{blue}][0]{.},
  morestring=[b]",
}

\newcommand{\CodeSymbol}[1]{\textcolor{magenta}{#1}}

\lstdefinelanguage{vrtx}
{
    morecomment=[s][\color{codegreen}]{<}{>},
    literate={\{}{{\CodeSymbol{\{}}}1
            {\}}{{\CodeSymbol{\}}}}1
            {:}{{\CodeSymbol{:}}}1
            {|}{{\CodeSymbol{|}}}1
}

\lstdefinelanguage{topsel}
{
    morecomment=[s][\color{codegreen}]{(}{)},
    literate={\{}{{\CodeSymbol{\{}}}1
            {\}}{{\CodeSymbol{\}}}}1
            {;}{{\CodeSymbol{;}}}1
}

\lstdefinelanguage{tapirconf}
{
  identifierstyle=\color{gray},
  morecomment=[f][\color{black}]{*\ tapir}
}

\newcommand{\file}[1]{{\color{green!50!blue} \texttt{#1}}}

\definecolor{codegreen}{rgb}{0,0.6,0}
\definecolor{codegray}{rgb}{0.5,0.5,0.5}
\definecolor{codepurple}{rgb}{0.58,0,0.82}
\definecolor{backcolour}{rgb}{0.95,0.95,0.92}

\lstdefinestyle{code}{
    backgroundcolor=\color{backcolour},   
    commentstyle=\color{codegray},
    keywordstyle=\color{magenta},
    stringstyle=\color{codepurple},
    basicstyle=\ttfamily\footnotesize,
    breakatwhitespace=false,         
    breaklines=true,                 
    captionpos=b,                    
    keepspaces=true,                 
    numbersep=5pt,                  
    showspaces=false,                
    showstringspaces=false,
    showtabs=false,                  
    tabsize=2,
    title=\lstname,
    belowskip=-1em,
    aboveskip = 1em
}

\newcommand\YAMLcolonstyle{\color{Maroon}\bfseries}
\newcommand\YAMLkeystyle{\ttfamily\footnotesize\color{black}\bfseries}
\newcommand\YAMLvaluestyle{\color{MidnightBlue}\bfseries}
\makeatletter

\newcommand\language@yaml{yaml}

\expandafter\expandafter\expandafter\lstdefinelanguage
\expandafter{\language@yaml}
{
  keywords={true,false,null,y,n,True,False,Null},
  keywordstyle=\color{ForestGreen}\bfseries,
  basicstyle=\YAMLkeystyle,                                 
  sensitive=false,
  comment=[l]{\#},
  morecomment=[s]{/*}{*/},
  commentstyle=\color{purple}\ttfamily,
  stringstyle=\YAMLvaluestyle\ttfamily,
  moredelim=**[il][\YAMLcolonstyle{:}\YAMLvaluestyle]{:},   
  moredelim=**[il][\YAMLvaluestyle{,}\YAMLvaluestyle]{,},   
  moredelim=**[il][\YAMLkeystyle{, }\YAMLkeystyle]{,\ },    
  moredelim=**[il][\YAMLkeystyle\texttt{\char125}\YAMLkeystyle]{\}}, 
  morestring=[b]',
  morestring=[b]",
  literate =    {---}{{\ProcessThreeDashes}}5
                {>}{{\textcolor{red}\textgreater}}1
                {|}{{\textcolor{red}\textbar}}1
                {\ -\ }{{\bfseries\ -\ }}3,
}

\lst@AddToHook{EveryLine}{\ifx\lst@language\language@yaml\YAMLkeystyle\fi}
\makeatother
\newcommand\ProcessThreeDashes{\llap{\ttfamily\footnotesize\color{MidnightBlue}\bfseries-{-}-}}

\newcounter{bla}

  \renewcommand{\equationautorefname}{Eq.}%

\newcommand{\Autoref}[1]{%
  \begingroup%
  \def\equationautorefname~##1\null{Eq.~(##1)\null}%
  \autoref{#1}%
  \endgroup%
}

\hypersetup{%
  colorlinks=true, linktocpage=true, pdfstartpage=1, pdfstartview=FitV,
  breaklinks=true, pageanchor=true,
  pdfpagemode=UseNone,
  plainpages=false, bookmarksnumbered, bookmarksopen=true, bookmarksopenlevel=1,
  hypertexnames=true, pdfhighlight=/O,
  urlcolor=RoyalBlue, linkcolor=Maroon, citecolor=RoyalBlue,
  pdftitle={tapir},
  pdfauthor={Marvin Gerlach, Florian Herren and Martin Lang},
  pdfsubject={},
  pdfkeywords={},
  pdfcreator={pdfLaTeX},
  pdfproducer={LaTeX with hyperref}
}

\titleformat*{\section}{\Large\bfseries}
\titleformat*{\subsection}{\large\bfseries}

\journal{Computer Physics Communications}

\begin{document}

\begin{frontmatter}
\title{\vskip-3cm{\baselineskip14pt
    \begin{flushleft}
      \normalsize TTP22-005\\ P3H-22-009\\ FERMILAB-PUB-22-025-T
  \end{flushleft}}
  \vskip1.5cm
  \tapir\\ A tool for topologies, amplitudes, partial fraction decomposition and input for reductions}

\author[a]{Marvin Gerlach}
\author[b]{Florian Herren\corref{author}}
\author[a]{Martin Lang}

\cortext[author] {Corresponding author.\\\textit{E-mail address:} florian.s.herren@gmail.com}
\address[a]{Institut f\"ur Theoretische Teilchenphysik\\ Karlsruhe Institute of Technology (KIT)\\ Wolfgang-Gaede-Stra\ss{}e 1, D-76128 Karlsruhe, Germany}
\address[b]{Fermilab, PO Box 500, Batavia, Illinois 60510, USA}

\begin{abstract}
The demand for precision predictions in the field of high energy physics has dramatically increased over recent years. Experiments conducted at the LHC, as well as precision measurements at the intensity frontier such as Belle II
require equally precise theoretical predictions to make full use of the acquired data. To match the experimental precision, second-, third- and, for certain quantities, even higher-order calculations in perturbative
quantum field theory are required.

To facilitate such calculations, computer software automating as many steps as possible is required. Yet, each calculation poses different challenges and thus, a high level of configurability is required.
In this context we present \tapir: a tool for identification, manipulation and minimization of Feynman integral families. It is designed to integrate in toolchains based on the computer algebra system \texttt{FORM},
the use of which is common practice in the field. \tapir can be used to reduce the complexity of multi-loop problems with cut-filters, topology mapping, partial fraction decomposition and alike.
\end{abstract}

\end{frontmatter}
\newpage
{\bf PROGRAM SUMMARY}

\begin{small}
\noindent
{\em Program Title:} \tapir                                       \\
{\em CPC Library link to program files:} (to be added by Technical Editor) \\
{\em Developer's repository link:} \href{https://gitlab.com/tapir-devs/tapir}{https://gitlab.com/tapir-devs/tapir} \\
{\em Code Ocean capsule:} (to be added by Technical Editor)\\
{\em Licensing provisions:} GPLv3  \\
{\em Programming language:} \python~$3$, \texttt{C++}                         \\
{\em Nature of problem:}\\
Multi-loop computations require the automatization of a large number of different tasks related to Feynman integral topologies. Among
them are the identification and minimization of integral topologies, partial fraction decomposition of topologies in the case of linearly dependent propagtors as well
as mapping scalar products of loop momenta to scalar functions.\\
{\em Solution method:}\\
The minimization of topologies is performed by comparison of their respective Nickel indices \cite{1}, even further minimization utilizes Pak's algorithm \cite{2}.
To efficiently map scalar products of loop momenta to scalar functions \texttt{FORM} \cite{3} code is generated.\\
{\em Additional comments including restrictions and unusual features:}\\
Minimization based on Pak's algorithm slows down for many lines and scales. A coarser minimization using the Nickel indices, however, is still possible.\\

\end{small}

\newpage

\setcounter{footnote}{0}

\section{\label{sec:intro}Introduction}

The evaluation of multi-loop scattering amplitudes is a challenging, yet indispensable, task to obtain precise predictions for scattering processes in high energy physics.
Precision measurements at the LHC require theoretical predictions for signal processes often at next-to-next-to-leading order in QCD,
see for example the \textit{precision Standard Model wishlist} \cite{Huss:2022ful}. Such precise predictions require the evaluation of two-loop amplitudes with up
to five external particles, out of which three can be massive (in the case of $t\overline{t}H$ production). Precision measurements at a future electron-positron collider operating at the $Z$-boson resonance will
require electroweak multiloop calculations at the two- and three-loop order, a task complicated by the multiple internal mass scales arising in such amplitudes \cite{Dubovyk:2022frj}.
On the contrary, for precision experiments in the flavour sector, state-of-the-art calculations do not deal with additional
scales or external particles but rather with even higher order QCD corrections. As an example, to match the current experimental precision, the inclusive semileptonic decay width of $B$ mesons
needs to be known at least at next-to-next-to-next-to-leading order in QCD, requiring a four-loop computation \cite{Fael:2020tow}.
In \cite{Heinrich:2020ybq}, the current frontiers in precision perturbative predictions for collider processes have been extensively marked out.

To obtain the scattering amplitude for a particular process, all contributing Feynman diagrams are generated based on a set of Feynman rules. Following diagram generation,
the numerator structure of each diagram is simplified before the resulting loop integrations are carried out.
The potentially large number of diagrams and the nontrivial numerator structures arising in gauge theories lead to a large number of individual Feynman integrals.
In general, it is not feasible to directly compute each individual integral. To overcome this issue, it is advantageous to group diagrams with similar underlying Feynman integrals by their topologies.
In the context of Feynman diagrams, topologies are also called \textit{integral families}.
Identifying a minimal set of topologies is crucial for higher loop calculations as the reduction of Feynman integrals to a set of basis functions, so-called \textit{master integrals},
is a tedious and expensive task in terms of computing time. Thus, duplicating work through a non-minimal set of topologies needs to be avoided.

Individual scalar expressions which appear in a typical $l$-loop amplitude can be expressed as a linear combination of Feynman integrals from different integral families.
Each family is characterized by a set of propagators and irreducible scalar products, $\{D_i\}$, composed of masses, loop momenta as well as external momenta.
Each Feynman integral is then specified by the family and a set of \textit{indices} $\{a_i\}$\footnote{In this paper, we always assume dimensional regularization with $d=4-2\epsilon$.}:
\begin{align}
  \label{eq:scalar-integral}
  I(a_1,...,a_N) = \int\!\!...\!\!\int \frac{\text{d}^d k_1 ... \text{d}^d k_l}{(2 \pi)^{ld}} \ \prod_{i=1}^N \frac{1}{D_i^{a_i}}\,.
\end{align}
In practical calculations, the form of \Autoref{eq:scalar-integral} can always be reached by proper use of projectors or other kinds of tensor reduction.
The scalar Feynman integrals can then be reduced to \textit{master integrals} using integration-by-parts relations \cite{Chetyrkin:1981qh,Tkachov:1981wb}. Subsequently,
the \textit{master integrals} need to be evaluated.

In calculations based on the method of reverse unitarity \cite{Anastasiou:2002yz} or in calculations comparing amplitudes in an effective and a full theory, special kinematics may often arise
and as a consequence give rise to linearly dependent propagators in Feynman integrals. As no unique set of integration-by-parts relations can be derived for families involving linearly dependent propagators,
new families with linearly independent propagators need to be derived by partial fraction decomposition.

Each of the aforementioned steps is a challenge in its own right and requires automatization for all but the most simple processes.
As a consequence, a large amount of programs have been developed that automatize one or more of these steps.

Feynman diagrams can be generated using \texttt{qgraf} \cite{Nogueira:1991ex} or \texttt{FeynArts} \cite{Hahn:1998yk,Hahn:2000kx}, while
their corresponding symbolic expressions can be simplified using programs such as \texttt{FeynCalc} \cite{Mertig:1990an,Shtabovenko:2016sxi,Shtabovenko:2020gxv}, \texttt{FormCalc} \cite{Hahn:2000kx}
or a plethora of non-public codes relying on computer algebra systems such as \texttt{Mathematica} or \texttt{FORM} \cite{Ruijl:2017dtg}.
The relevant Feynman rules for a given Lagrangian can be automatically obtained using \texttt{FeynRules} \cite{Christensen:2009jx,Alloul:2013bka} which supports a variety of output
formats, such as \texttt{UFO} \cite{Degrande:2011ua}.

In the case of one-loop calculations, the Passarino-Veltman method \cite{Passarino:1978jh} allows to express any one-loop diagram with propagators that depend quadratically on the momentum
as a linear combination of a limited set of basis functions, allowing to completely automate the symbolic computation of one-loop diagrams. This algorithm is implemented in programs and libraries such as
\texttt{FeynCalc}, \texttt{FormCalc}, \texttt{LoopTools} \cite{Hahn:1998yk}, \texttt{OneLOop} \cite{vanHameren:2010cp}, \texttt{Package-X} \cite{Patel:2015tea}, \texttt{HEPMath} \cite{Wiebusch:2014qba}, \texttt{Collier} \cite{Denner:2016kdg}, \texttt{QCDLoop} \cite{Carrazza:2016gav} or \texttt{NLOX} \cite{Honeywell:2018fcl,Figueroa:2021txg}. In addition to the Passarino-Veltman method, most of these packages also provide functionality for
evaluating the resulting basis functions numerically. Furthermore, programs such as \texttt{Madgraph} \cite{Alwall:2014hca},
\texttt{Grace} \cite{Belanger:2003sd} or \texttt{GoSam} \cite{Cullen:2011ac,Cullen:2014yla} allow for completely automated computation of one-loop amplitudes.
 
As such an algorithm is not known at two loops and beyond, dedicated tools are required for the different parts of such a calculation.
The generation of symbolic expressions for diagrams, identification and minimization of topologies, as well as the rewriting of propagators and scalar products in terms of scalar functions, as defined in
\Autoref{eq:scalar-integral}, have been implemented in programs such as \texttt{DIANA} \cite{Tentyukov:1999is}, \texttt{q2e} and \texttt{exp} \cite{Harlander:1998cmq,Seidensticker:1999bb,q2eexp}, \texttt{TopoID} \cite{Hoff:2016pot},
\texttt{ALibrary} \cite{alibrary}, \texttt{Feynson} \cite{feynson} or \texttt{LIMIT} \cite{Herren:2020ccq}. Partial fraction decomposition of linearly dependent integrands can be performed
by \texttt{TopoID}, \texttt{APart} \cite{Feng:2012iq}, \texttt{Multivariate Apart} \cite{Heller:2021qkz} and \texttt{LIMIT}.

Integration-by-parts relations can be solved in a symbolic manner for certain integral families, either by hand or by programs such as \texttt{LiteRed} \cite{Lee:2013mka}.
Prominent examples are massive tadpole or massless two-point integrals. Both types of integrals have been implemented in the dedicated computer programs
\texttt{MINCER} \cite{Larin:1991fz}, \texttt{MATAD} \cite{Steinhauser:2000ry}, \texttt{FORCER} \cite{Ruijl:2017cxj} and \texttt{FMFT} \cite{Pikelner:2017tgv}.
For arbitrary families, where no such algorithm is known, programs including \texttt{FIRE} \cite{Smirnov:2008iw,Smirnov:2019qkx}, \texttt{Reduze} \cite{Studerus:2009ye,vonManteuffel:2012np} or
\texttt{Kira} \cite{Maierhofer:2017gsa,Klappert:2020nbg} generate systems of linear equations by evaluating integration-by-parts relations for fixed indices.

While a number of programs implementing different parts of multi-loop calculations exist, most of them only implement one specific task.
As topology identification and manipulation, partial fraction decomposition and the generation of symbolic expressions all rely on the underlying graphs
associated with the Feynman diagrams under consideration, combining these tasks in a single program allows for a unified treatment without interoperability issues
that can arise when combining several individual programs.

With this in mind, we introduce \tapir: a tool for topologies, amplitudes, partial fraction decomposition and input for reductions.
\tapir is written in order to replace two of the aforementioned programs and expand on their scope: \texttt{q2e} and \texttt{LIMIT}.
\texttt{q2e} is written to generate \texttt{FORM} expressions for Feynman diagrams generated by \texttt{qgraf} and providing information about their graph structure to \texttt{exp}.
As a consequence, \tapir inherits input formats for Feynman diagrams and Feynman rules from \texttt{q2e} while generating compatible output files.
In contrast to \texttt{q2e}, \texttt{LIMIT} works at the level of Feynman integrals, allowing for partial fraction decomposition and minimization of integral families.
\tapir can use the information about the graphs associated to the Feynman diagrams for a given problem to not only generate symbolic expressions but also to generate \texttt{FORM} code to perform
the same tasks \texttt{LIMIT} performs.

\begin{figure}
  \centering
  \includegraphics[width=1\textwidth]{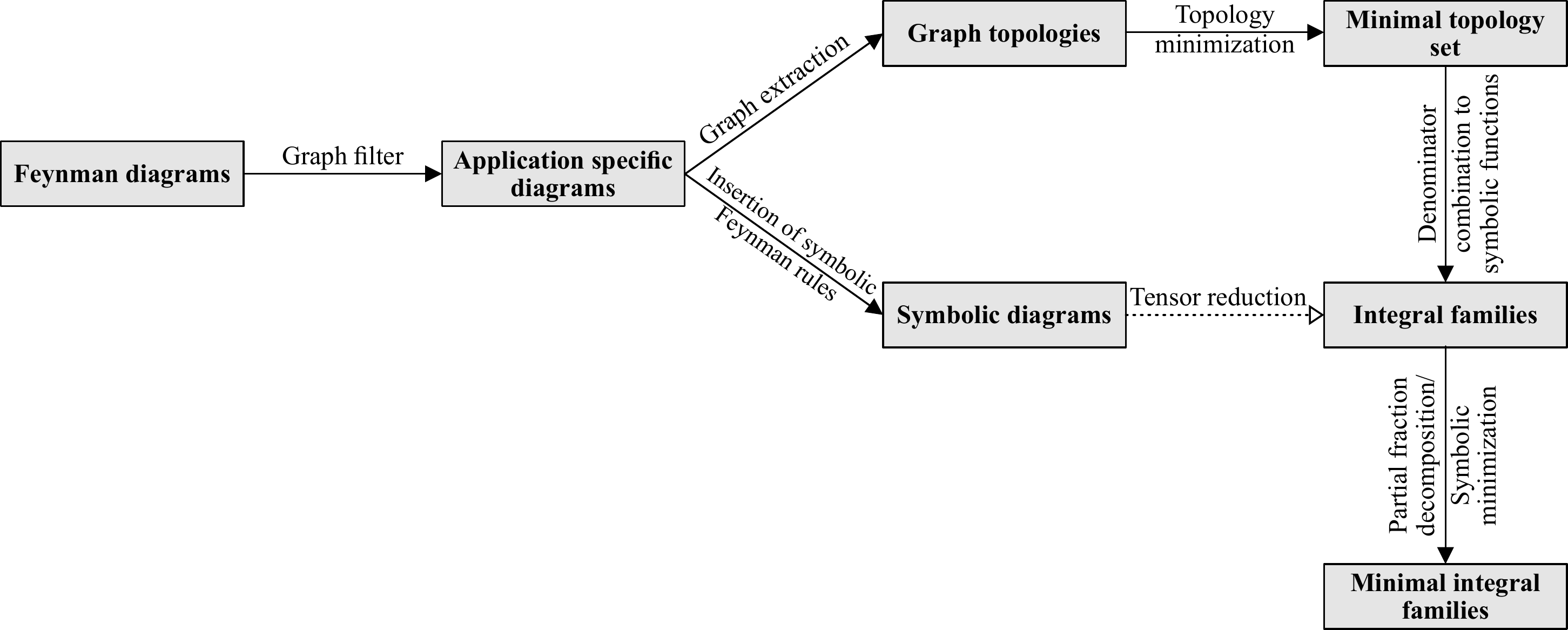}
  \caption{The application of \tapir is manifold, although the common workflow is oriented on well-established \texttt{FORM}-based setups for multi-loop computations. Tensor reduction is not part of \tapir and must be applied externally.}
  \label{fig:workflow}
\end{figure}

\tapir is written in \python~$3$ and is developed in a test-driven, as well as self-documented style to ensure reliability and maintainability.
The workflow of \tapir is illustrated in \autoref{fig:workflow}.

In a first step, diagrams are read from the output of \texttt{qgraf}, the so-called \file{qlist} file. Next, diagram filters, in addition to the \texttt{qgraf} built-in filters,
can be used to extract the diagram classes of interest. Of particular interest is the \textit{cut filter}, based on an algorithm discussed in Ref.~\cite{Hoff:2015kub},
which removes diagrams with unwanted Cutkosky cuts.
Diagram filters act in general non-destructively, i.e.\ they only include or exclude diagram classes with specific features, but do not alter the diagrams themselves.
The symbolic diagram representation is then built from a set of Feynman rules which are defined in \file{vrtx} files for interactions and \file{prop} files for propagators.
Symbolic expressions in \tapir are represented in syntax for the computer algebra system \texttt{FORM} and written to so-called \file{dia} files. Note that these files are
compatible with \texttt{q2e} input and output files and we also use the same naming scheme. Alternatively, Feynman rules can be read from \texttt{UFO} models generated
by \texttt{FeynRules}, allowing for the automated generation of Feynman rules for many different kinds of quantum field theory models.

In addition, \tapir offers destructive topology filters which change the topology structure. An example use case is the removal of auxiliary particle lines from the topology,
whose propagators do not transfer momentum.

The graph information of the diagrams is later used to identify the topologies in which the propagators, also called denominator functions, $\{D_i\}$ resemble the edges of the underlying graph.
Other objects such as numerators or eikonal propagators cannot be expressed in this graph representation. We can only account for them as symbolic objects of the scalar integral family function.

For every graph, the so-called \textit{Nickel index} \cite{Nickel:1977} is then computed. It is a unique naming convention which can be used to identify similar graphs.
The Nickel index is used to minimize the graph topologies by building a hash-table with the index as hashing function. Together with the information on how the graph edges must be named to
be canonically ordered with respect to the index, we can also give a mapping description between line momenta of similar topologies.

Instead of using the Nickel index for momentum mapping, the program \texttt{exp} can be used. It has the advantage of finding mappings on (sub-)topologies rather quickly for a small amount of target topologies.
For larger problems, on the other hand, matching Nickel indices has a better algorithmic scaling and enables minimization of topologies in a highly parallelizable manner.
For the usage with \texttt{exp}, an \file{edia} file as well as a \file{topsel} file can be generated which include
the topological information of the diagrams. Also these file formats were adopted from \texttt{q2e} and \texttt{exp}.

The remainder of this article is structured as follows: in \autoref{sec:concepts} we introduce the key ideas and algorithms implemented in \tapir. In \autoref{sec:definitions} we discuss
the format of the input and output files, while in \autoref{sec:usage} we present explicit examples for the usage of \tapir. We summarize the features of \tapir in \autoref{sec:sum}.
Finally, \ref{sec:UFOReader} provides details on the interface between \tapir and \texttt{UFO} models, while \ref{sec:yaml} introduces the \texttt{YAML} input file format.

\section{\label{sec:concepts} Key concepts and algorithms}
In \tapir we incorporated a few noteworthy algorithms and ideas, summarized in this section.

\subsection{\label{sec:nickel}Nickel index}
To compare different Feynman graph topologies the use of a canonical graph naming scheme is suited best. This has the advantage of efficient pre-computation which can be highly parallelized. The canonical label can afterwards be used as a unique identifier to quickly find graph topologies. Such labels are, for example, implemented in the program pair \texttt{nauty} and \texttt{Traces} \cite{MCKAY201494}. Although these programs are publicly available and widely used, their graph labeling for our purposes\footnote{In the language of graph theory Feynman graphs are edge-colored graphs with multiple edges and self-loops.} is not directly applicable without additional effort.

To have control over the labeling algorithm we decided to implement a version of the \textit{Nickel index} as a canonical graph label. The algorithm for its computation is described in detail in Ref.~\cite{Batkovich:2014bla}.

\begin{figure}
  \centering
  \includegraphics[height=3cm]{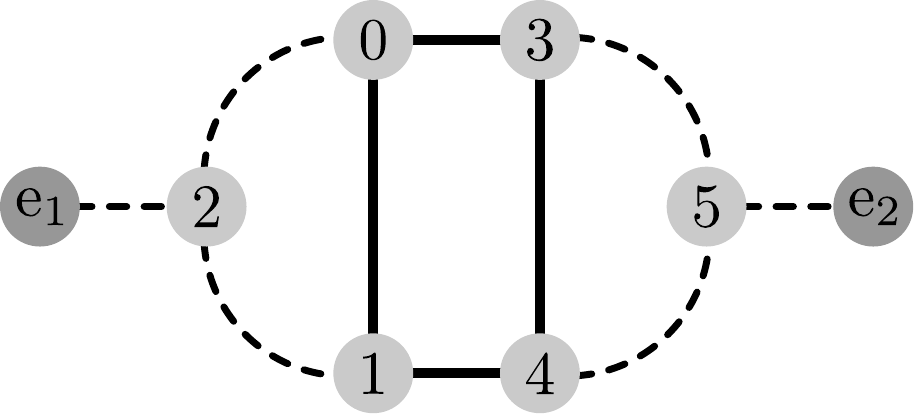}
  \caption{The Nickel index gives a canonical labeling of the vertices of a Feynman graph. The dashed lines denote massless propagators and solid lines propagators with mass \texttt{M1}.}
  \label{fig:nickel-example}
\end{figure}

An example topology is given in \autoref{fig:nickel-example}. Its Nickel index is given by

\begin{equation*}
  \texttt{123|24|e|45|5|e|~:~M1\_\_M1|\_M1|q1|M1\_||q2|}\,
\end{equation*}

and consists of two parts. The first part uniquely describes how the vertices are connected. The information for each vertex is written as the numeric labels of its adjacent vertices, but in such a way that only the vertex with the smaller number provides this information.

For example, the first entry \texttt{123} describes vertex 0. It is connected to vertices 1, 2 and 3. Vertex 1 is connected to 0, 2 and 4. But since the entry for vertex 0 already described the former connection we skip it and get the entry \texttt{24} for vertex 1. 

Connections to external lines are denoted by ``\texttt{e}''. The different vertex entries are separated by a vertical line ``\texttt{|}''. Thus self-loops and multiple edges can be easily represented.

The second part of the Nickel index describes the edge coloring. In the case of Feynman diagram topologies, this is either the mass of the propagator for internal lines or the momentum for external lines. For massless internal lines the entry is kept empty. The order is the same as in the first part, but the individual edge entries are separated by an underscore ``\texttt{\_}''. This notation has the advantage that topologies with special kinematics can be identified, e.g.\ when two external momenta are equal.

Vertex 0 in this example is connected to vertices 1 and 3 by a line with mass \texttt{M1} and to vertex 2 by a massless line. The corresponding second part of the Nickel index to vertex 0 is thus \texttt{M1\_\_M1}.

Although the vertex numbering is not unique, the Nickel index is.
To obtain a unique label, all possible \textit{Nickel notations} according to different vertex enumerations have to be computed. This notation is simply a string combination of the two described parts. Then all notations are compared and the one with the lowest lexicographical order is kept as the Nickel index. Hence, the full algorithm has at least a complexity of $\mathcal{O}(v!)$, where $v$ is the number of vertices.

Our implementation of the algorithm as a \texttt{C++} sub-module provides additional information how the edges are ordered in the Nickel index. We can use this information to find mappings of line momenta of graphs with the same topologies.

A drawback of this approach is its inefficiency in mapping topologies with fewer lines on larger topologies. For this, in principle, all possible Nickel indices need to be determined for the possible substructures of the larger topology with certain lines contracted to a point, since they cannot be inferred from the Nickel index of the parent topology. To overcome this problem, the program \texttt{exp} or Pak’s algorithm, described in the next section, can be used.

\subsection{Pak's algorithm}
With the method described above only graphs can be compared. A more general procedure to compare different integral families with or without graph representation is provided by an algorithm by A. Pak \cite{Pak:2011xt}.

To make use of it, \Autoref{eq:scalar-integral} is written as a Feynman parameter integral. Integration over loop momenta finally leads to
{\small
\begin{equation}
  \label{eq:feynman-param}
  \begin{aligned}
    I(a_1,...,a_N) = \ \frac{i^l}{(4\pi)^{dl/2}} \frac{\Gamma\left(\underset{i}{\sum} a_i - \frac{ld}{2}\right)}{\underset{i}{\prod}\Gamma(a_i)} \int\displaylimits_0^{\infty}\!...\!\int\displaylimits_0^{\infty}  \left(\prod_i \text{d} x_i \ x_i^{a_i-1}\right) \delta\left(\sum_i x_i-1\right) \frac{\mathcal{U}^{\underset{i}{\sum} a_i - \frac{(l+1)d}{2}}}{\mathcal{F}^{\underset{i}{\sum} a_i - \frac{ld}{2}}}\,.
  \end{aligned}
\end{equation}
}

The so-called \textit{Symanzik polynomials} $\mathcal{U}$ and $\mathcal{F}$ can be computed in many different ways (see Ref.~\cite{Bogner:2010kv} for an elaborate overview).
We found the most efficient way to be the method implemented in Ref.~\cite{Asmirnov:uf}.

The idea of Pak's algorithm is to find a canonical enumeration of Feynman parameters such that the monomials of $\mathcal{U} \cdot \mathcal{F}$ are lexicographically maximal. This is done by iterative renaming of the $x_i$ to find the lexicographically maximal name for $x_1$. After that follows the search for the maximal $x_2$ with $x_1$ fixed, and so on. 

As an example we start with the simple integral family

\begin{equation}
  I(a_1, a_2) = \int \frac{\text{d}^d k}{(2 \pi)^{d}} \ \frac{1}{(k^2)^{a_1} \ ((k+q)^2-m^2)^{a_2}}\,.
\end{equation}

The Symanzik polynomials entering \Autoref{eq:feynman-param} for this family are given by

\begin{equation}
  \begin{aligned}
    \mathcal{U} &= x_1 + x_2\,,\\
    \mathcal{F} &= x_2^2 m^2 + x_1 x_2 (m^2 - q^2)\,.\\
  \end{aligned}
\end{equation}

$x_1$ and $x_2$ can be renamed at will. For a canonical labeling we have to compare the follwing two possibilities lexicographically:

\begin{equation}
  \begin{aligned}
    \mathcal{U \cdot F} \ &\overset{\text{lex.}}{>} \ (\mathcal{U \cdot F})_{x_1 \leftrightarrow x_2} \\
    x_1^2 x_2 (m^2-q^2) + x_1 x_2^2 (2 m^2 - q^2) + x_2^3 m^2 &\overset{\text{lex.}}{>}x_1^3 m^2 + x_1^2 x_2 (2 m^2 - q^2) + x_1 x_2^2 (m^2-q^2) 
  \end{aligned}
\end{equation}

The lexicographic comparison can be a simple string comparison of the individual (properly ordered) monomials. The global definition of ``maximal'' is irrelevant in this context, since only the relative ordering matters when comparing integral families.

The algorithm starts with expressing the polynomial $\mathcal{U} \cdot \mathcal{F}$ as a matrix, with rows representing vectors of the powers of the $x_i$ for each monomial. Then, the maximal labeling according to $x_1$ is found by iterative switching of the first with the \textit{i}'th column. The matrix rows are sorted according to their entries, and the variant with the lexicographically maximal first column vector is kept. Then the same procedure is applied to the second column, keeping $x_1$ fixed. This continues until all $x_i$ names are determined.

For this example the matrix is given by

\begin{equation}
  \begin{aligned}
    \begin{pmatrix}
      2 & 1 & m^2-q^2\\
      1 & 2 & 2m^2-q^2\\
      0 & 3 & m^2\\
    \end{pmatrix}\,.
    \end{aligned}
\end{equation}

An additional column is added to take also the prefactors into account. The renaming of $x_1 \leftrightarrow x_2$ now gives the following column-sorted matrix:

\begin{equation}
  \begin{aligned}
    \begin{pmatrix}
      3 & 0 & m^2\\
      2 & 1 & 2m^2-q^2\\
      1 & 2 & m^2-q^2\\
    \end{pmatrix}\,.
    \end{aligned}
\end{equation}

Comparing the vectors of the first columns, we see that the renaming leads to a lexicographically larger result. 

The formal complexity of this procedure grows at least as $\mathcal{O}(e!)$, where $e$ is the number of propagators. If at some point two or more Feynman parameters acquire the same (maximal) lexicographic order, a symmetry is found. When comparing $\mathcal{U} \cdot \mathcal{F}$ of different integral families, one has to iterate over all symmetries and compare each possible $x_i$ enumeration individually. This slows the procedure down and can only be parallelized in exchange for an increased memory usage. 

In the original definition of the algorithm, copies of the matrix are created if a symmetry appears. Then the algorithm continues for all subsequent matrices. Keeping only a single matrix while memorizing the symmetries is also known as the \textit{Light Pak algorithm}~\cite{Heinrich:2021dbf}.

Pak's algorithm is nevertheless an outstanding tool for our purposes, as we use it to compare integral families after partial fraction decomposition, which do not necessarily possess a graph representation.

\subsection{Partial fraction decomposition}
In general, the denominator functions $\{D_i\}$ in an integral family can be linearly dependent. Thus, some sort of partial fraction decomposition must be performed before the reduction with programs such as \texttt{FIRE} is possible. 

For a generic partial fraction decomposition, we employ the idea of Ref.~\cite{Hoff:2015kub}, where the problem was reformulated to finding an appropriate Gröbner basis \cite{Buchberger:phd}. 
The method starts by identifying linear relations between propagators. The easiest way to construct these relations is by representing each denominator function in terms of loop momenta $k_i$, external momenta $q_i$ and masses $m_i$. For this purpose we construct a tuple of scalar products and masses that appear in the denominator functions $D_i$. The ordering in the tuple is chosen such that terms with and without loop momenta are split, e.g.

\begin{align}
  \mathbf{P} = \begin{pmatrix}
    k_1^2\\
    k_1\cdot q_1\\
    k_1\cdot q_2\\
    \hline
    q_1^2\\
    q_2^2\\
    q_1\cdot q_2\\
    m^2
  \end{pmatrix}  \equiv \begin{pmatrix}
    \mathbf{P_l} \\
    \mathbf{P_n}
  \end{pmatrix}\,.
\end{align}

Then a coefficient matrix $M$ is constructed which expresses the inverse denominator functions linearly in terms of $\mathbf{P}$:

\begin{align}
  M \cdot \mathbf{P} = \begin{pmatrix}
    D_1^{-1}\\ 
    D_2^{-1}\\ 
    D_3^{-1}\\
    \vdots\\
    D_N^{-1}\\
  \end{pmatrix} \equiv \mathbf{\tilde{D}}\,.
\end{align}

The linear relations between the $D_i^{-1}$ are found by computing a left-inverse $M_L^{+}$ of $M$ and performing Gaussian reduction such that $M_L^{+}$ takes a row echelon form. This is easily achieved by transforming the matrix

\begin{align}
  \left(M \ \vline \ \text{diag}\left(\mathbf{\tilde{D}}\right) \right)
\end{align}

to row echelon form. The rows containing only zeros in the $M$-side of this matrix lead to linear equations of the form
\begin{align}
  \label{eq:zero-partfrac}
  0 = \sum_i a_{ji} D_i^{-1}\,,
\end{align}
whereas rows containing only non-zero entries of $\mathbf{P_n}$ (i.e.\ loop-momenta independent) give rise to linear equations of the form
\begin{align}
  \label{eq:nonzero-partfrac}
  b_j = \sum_i a_{ji} D_i^{-1}\,.
\end{align}

Both, \Autoref{eq:zero-partfrac} as well as \Autoref{eq:nonzero-partfrac}, can be used iteratively to reduce the number of denominator functions of an integral family. Gröbner bases can help to apply these relations in such a way that only families with linear independent denominator functions remain. Thus, we find the partial fraction reduction steps directly, when we use the following polynomial basis as input:

\begin{align}
  K = \left\{\left\{\sum_i a_{ji} D_i^{-1}\right\}, \ \left\{\sum_i a_{ki} D_i^{-1}-b_k\right\}, \ \left\{D_l\cdot D_l^{-1}-1\right\} \ \vline \ \forall \ j,k,l \right\}\,.
\end{align}

Applying the Buchberger algorithm with respect to all denominators and inverse denominators $\{D_i, D_i^{-1} \}$ gives the iterative (and always deterministic) prescription for partial fraction decomposition with arbitrary indices for a given integral family we were looking for. The Buchberger algorithm was found together with the idea of Gröbner bases in Ref.~\cite{Buchberger:phd}.

The flavor of the actual implemented Buchberger algorithm defines how it scales with the number of linearly dependent denominator functions. However, the worst possible scaling behavior is double exponential \cite{MAYR1982305}. At present, different approaches using heuristics and advanced monomial orderings are widely available in many computer algebra systems. As for all symbolic manipulations within \tapir, we use \texttt{sympy}~\cite{10.7717/peerj-cs.103} for this purpose with its implementation of the \textit{improved Buchberger algorithm}~\cite{doi:10.1137/1036089}.

\subsection{\label{sec:cutkosky}Cutkosky cut filter}
In some scattering problems in quantum field theory it is useful to utilize the unitarity constraint to apply the optical theorem. With it, it is possible to relate quantities such as total cross sections or decay rates to the imaginary part of a scattering amplitude. To compute the latter, it is common to use the \textit{Cutkosky cutting rules}~\cite{Cutkosky:1960sp}, which state that the imaginary part of a Feynman diagram equals the sum of all subsequent ``cut'' diagrams. A cut is defined as a set of loop-propagators which separates external vertices when removed. These cut propagators are then treated as on-shell.
The Cutkosky rules allow to exclude diagrams that only contain cuts leading to final states that are not allowed kinematically. In addition, diagrams with specific cuts, e.g.\ corresponding to a specific decay mode, can be singled out. 

For these purposes \tapir offers a \textit{cut filter} option to collect only diagrams with certain cut properties. It allows to effectively reduce the number of diagrams in an early calculational stage, and hence to reduce the complexity of problems in which decay processes play a crucial role.

The cut filter is based on an algorithm by A. Pak, described in great detail in Ref.~\cite{Hoff:2015kub}. Several instructions need to be provided to the algorithm, such as whether a filter shall be applied inclusively or exclusively. To find cuts the points through which the external momenta enter need to be divided into ``sources'' and ``sinks'',
such that a cut is defined as a cut through non-bridge lines (i.e.\ lines that are part of loops) that separates sinks from sources.
Furthermore, the types and numbers of particles that can be cut for a kinematically allowed configuration need to be specified. Several cut filter rules can be applied simultaneously to enable complex filter combinations.

The mentioned algorithm to find the diagrams that are allowed according to the given rules, can be described as follows:

\begin{itemize}
  \item[0)] Invert all disallowed ranges of cut particles and reformulate them as an allowing rule, e.g.\ instead of disallowing cuts with two or three particles allow cuts with zero to one and four and more particles.
  \item[1)] Categorize, a priori, all lines of a diagram either in the groups \texttt{M} (must be cut), \texttt{C} (can be cut) or \texttt{N} (must not be cut) according to the filter arguments.
  \item[2)] Colorize all source vertices identically (label \texttt{0}), as well as all sink vertices (label \texttt{1}).
  \item[3)] Specify the colors of all other internal vertices by the following rules as far as possible: Adjacent vertices that are separated by an \texttt{N} line have the same color; the ones connected by an \texttt{M} line must have different colors. Introduce new ancillary colors if the coloring is ambiguous.
  \item[4)] Iterate over all possible assignments of the newly introduced colors to the color of the sources (\texttt{0}) or of the sinks (\texttt{1}) respecting the rules of step 3 and the defined filter restrictions in every iterative step.
  \item[5)] Successful bi-coloring of a graph indicates that the separation between sources and sinks is allowed and the diagram is kept. On the other hand, diagrams for which no bi-coloring can be applied are dropped.
\end{itemize}

Step 0 must be applied only once, whereas the rest is done for each diagram individually. Splitting the vertex coloring procedure into different steps allows dropping diagrams already in step 3 if vertex clusters cannot be sufficiently separated. 

Due to this heuristic exclusion approach, the algorithm does not suffer as much from combinatorially expensive recoloring as the naive brute force approach would. Hence, it is well suited for large scale problems.

The cut filter has been tested in the case of real-virtual and double-real corrections to Higgs boson pair production. These corrections were calculated in Refs.~\cite{Davies:2019xzc,Davies:2021kex} using the
program \texttt{gen} \cite{Pak:unpublished} for cut filtering.

\section{\label{sec:definitions}Definitions}
To use \tapir several file formats need to be introduced. The \file{config}, \file{prop} and \file{vrtx} files provide options and Feynman rules to \tapir and need to be specified by the user.
All other files discussed in this section provide additional information about the analyzed diagrams and topologies, or have specific formats to be directly read from \texttt{FORM} or \texttt{Mathematica} programs. Most of the files described here follow the conventions and implementations of \texttt{q2e} and \texttt{exp} to ensure compatibility with existing setups and workflows.

In future releases, additional files and options might be introduced. Hence, we always refer to the project repository and the user documentation therein as a single source of reference.

\subsection{\file{config} files}
To hand complex instructions to \tapir, the command line alone is not sufficient due to the vast set of options. Instead, a \file{config} file has to be provided with instructions of the following form:

\begin{lstlisting}[language=tapirconf]
* tapir.[option keyword] {: [option argument 1]} {: [option argument 2]} ...
\end{lstlisting}

Lines without the \texttt{* tapir} tag at the beginning are treated as comments. Also note the mandatory whitespace after the asterisk.

To use \tapir in a pipeline setup, i.e.\ different program calls from the same directory, it is often useful to write multiple \file{config} files and provide each of them separately as a command line argument. In addition, \tapir has further command line options which can be provided via the \file{config} file as well, with prioritization of the former.

The following example instructs \tapir to read a \texttt{qgraf} output file \file{qlist.2} and draw some diagrams thereof with representative topologies:

\begin{lstlisting}[language=tapirconf]
Propagator and vertex Feynman rules
* tapir.propagator_file qcd.prop
* tapir.vertex_file qcd.vrtx

qgraf input
* tapir.qlist qlist.2

Output
* tapir.repart representatives.tex

Define mass of top quark
* tapir.mass t : M1

Drawing options
* tapir.draw_particle t : fermion : $t$
* tapir.draw_particle g : gluon : $g$
* tapir.draw_particle h : scalar : $h$
\end{lstlisting}

The output \texttt{representatives.tex} is a \LaTeX \ file using the \texttt{TikZ-Feynman} package~\cite{Ellis:2016jkw} to draw the corresponding diagrams. It can be compiled using the broadly available program \texttt{lualatex}. The drawing options at the end specify how the propagators of the individual particles are drawn, and what label should be printed next to them. The Feynman rule files \file{qcd.prop} and \file{qcd.vrtx} are mandatory because they include the particle and anti-particle information of the theory.

\subsection{\file{prop} and \file{vrtx} files}\label{subsec:propvrtx}
The \file{prop} and \file{vrtx} files define the Feynman rules of the propagators and vertices that are declared in \qgraf{} \file{lagrangian} files.
They contain symbolic \FORM code for each tuple of particles that correspond to the specified Feynman rule, which can be used to produce \file{dia} (see \autoref{subsec:diafiles}) files from \qgraf's output \file{qlist} file.
It is possible to have multiple entries for the same tuple of particles to separate different contributions to a given Feynman rule, for example in theories with multiple four-fermion operators defined by the same set of particles.
Entries in the \file{vrtx} or \file{prop} files with the same particle content are added up at the amplitude level due to separation into individual diagrams. 
All entries of \file{prop} and \file{vrtx} files are of the same generic form:
\begin{lstlisting}[language=vrtx]
{particle1,particle2(,particle3,...):Lorentz|QCD|QED|EW}
\end{lstlisting}
In the case of propagators, only \texttt{particle1} and \texttt{particle2} are used, vertices contain at least three particles. The particles \texttt{particle1} and \texttt{particle2} which share a propagator are interpreted as anti-particles of each other.
The particle content is separated from the Feynman rule by a ``\texttt{:}'', and the Feynman rule itself is divided into four different folds, separated by ``\texttt{|}''.
The \texttt{QED} and \texttt{EW} folds are usually empty, as they are historical structures that do not serve any special purpose anymore.
The \texttt{Lorentz} and \texttt{QCD} folds serve as a possibility for the user to isolate different parts of the calculation, in particular the color algebra and the Dirac structure, and tackle them separately, which is particularly useful for large-scale calculations.
Of course, the decision whether to split the Feynman rules or not is entirely up to the user, in particular the \texttt{QED} and \texttt{EW} folds can be used to further split a calculation into smaller pieces. 
In most cases, the Feynman rules involve template variables such as \texttt{<lorentz\_index\_particle\_3>}, which refers to the Lorentz index of the third particle entering the vertex, that will be replaced by a definite index or symbol once \tapir generates the \file{dia} file.
These placeholder variables can be used to construct products of tensors in different spaces, e.g.\ for spinor and color structures.
The entry of each fold of a Feynman rule must begin with a multiplication asterisk ``\texttt{*}'', as the symbolic diagram expression results from concatenation (i.e.\ multiplication) of several strings from vertices and propagators.\newline
Examples of typical \file{prop} and \file{vrtx} entries are given in \autoref{subsec:threegluonex} and \ref{sec:UFOReader}.

\subsection{\file{dia} files}\label{subsec:diafiles}

\begin{figure}
  \centering
  \includegraphics[height=7cm]{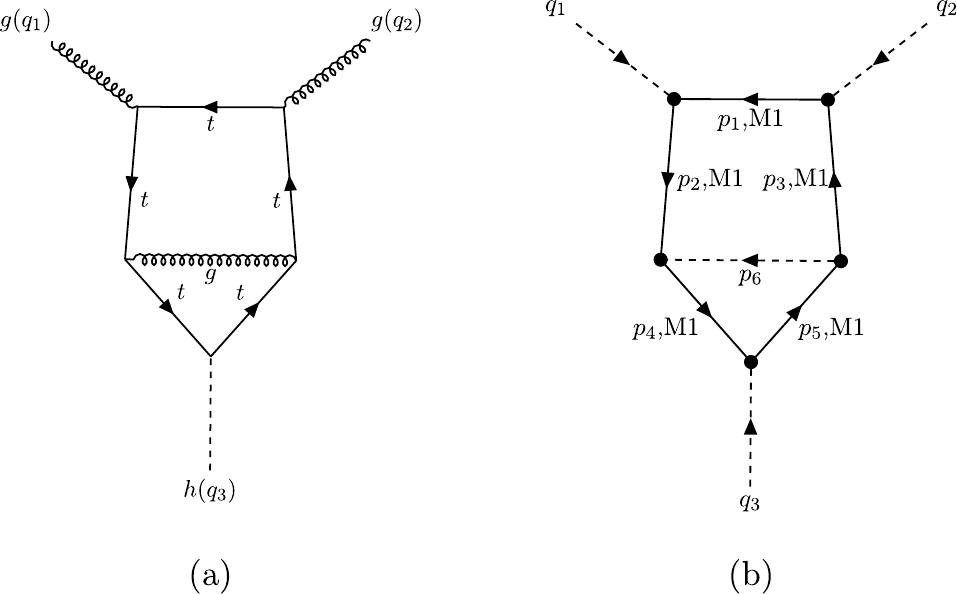}
  \caption{The graphic representations of the diagram (a) and topology (b) of the examples in \autoref{sec:definitions} were generated using the \texttt{--diagramart} and \texttt{--topologyart} command line options. The line labels were shifted for better readability. Note: external momenta are always treated as incoming.}
  \label{fig:ggh2l}
\end{figure}

Based on the Feynman rules given in \file{prop} and \file{vrtx} files \tapir generates four \FORM code folds
for each diagram: \texttt{Lorentz}, \texttt{QCD}, \texttt{QED} and \texttt{EW} fold.
These folds are written to the \file{dia} file. Note that the \file{dia} file is compatible with \file{dia} files produced by \texttt{q2e} and can serve as an input to \texttt{exp}.

As an example, consider the two-loop diagram of \autoref{fig:ggh2l} (a) contributing to Higgs boson production in gluon fusion:
\begin{lstlisting}[language=form]
*--#[ d2l3 :

	(-1)*1*nh
  *FT1(mu2)
  *FT1(p3)
  *FT1(nu17)
  *FT1(p5)
  *FT1(M1)
  *FT1(p4)
  *FT1(nu18)
  *FT1(p2)
  *FT1(mu1)
  *FT1(p1)
  *Dg(nu17,nu18,p6)
  ;

	#define TOPOLOGY "arb"
	#define INT1 "arb"

*--#] d2l3 :

*--#[ fqcd2l3 :

  1
  *GM(a(2),j7,j12)
  *d_(j12,j11)
  *GM(b(17),j11,j16)
  *d_(j16,j15)
  *d_(j15,j14)
  *d_(j14,j13)
  *GM(b(18),j13,j10)
  *d_(j10,j9)
  *GM(a(1),j9,j8)
  *d_(j8,j7)
  *prop(b(17),b(18))
  ;

*--#] fqcd2l3 :

*--#[ fqed2l3 :
	1
*--#] fqed2l3 :

*--#[ few2l3 :
	1
*--#] few2l3 :
\end{lstlisting}

The two numbers in the name of the \texttt{Lorentz} fold denote the number of loops and the number of the diagram, respectively.
All other folds follow this pattern, but have additional prefixes to denote the respective fold. Each fold can be loaded using \texttt{FORM}'s \texttt{\#include} preprocessor directive.

The first line in the \texttt{Lorentz} fold contains the symmetry factor provided by \texttt{qgraf} (\texttt{(-1)} in this case),
factors for counting closed fermion loops as defined in the \file{config} file\footnote{To obtain information about closed fermion loops, the option \texttt{tapir.contract\_fermion\_lines} must be used and all fermion names must start with ``f'' to indicate them as such. With this option it is not possible to directly define e.g.\ four-fermion interactions.} (\texttt{*nh} in this case) and a \texttt{*1} in the case of the previous factors being trivial and thus absent. 
The following lines contain the Feynman rules for all occuring propagators and vertices, filled with the respective open indices (e.g.\ \texttt{mu2}), line momenta (e.g.\ \texttt{p1}) and masses (e.g.\ \texttt{M1}). The demonstrated Feynman rules incorporate a non-commuting object \texttt{FT1} which corresponds to a Dirac gamma matrix, either with an open index or contracted with a line momentum.
Similarly, the \texttt{QCD} fold contains the color part of the Feynman rules, in which \texttt{GM} refers here to the Gell-Mann matrix.

In this example the other two folds are empty, but they can be populated depending on the Feynman rules.

\subsection{\file{edia} and \file{topsel} files}\label{subsec:ediafiles}
The information necessary for topology identification of each diagram is contained in the \file{edia} file.
Each entry takes the form
\begin{lstlisting}[language=topsel]
{Name; Number of lines; Number of loops; Number of independent ext. legs; 
  Number of masses; List of scales; List of line momenta}
\end{lstlisting}
where the name is the same as the name of the \texttt{Lorentz} fold in the \file{dia} file, the list of scales is taken from the \file{config} file and
the list of line momenta contains entries of the form \texttt{(Momentum,Mass:v1,v2)} where \texttt{v1} and \texttt{v2} are the vertices the momentum flows in between.
For external momenta, the two numbers \texttt{v1} and \texttt{v2} denote where the momentum enters and leaves the graph, respectively.

In the case of the diagram ``2l3'' of the previous section, the corresponding topology is visualized in \autoref{fig:ggh2l}~(b). Its \texttt{edia} entry takes the following form:
\begin{lstlisting}[language=topsel]
{d2l3;6;2;2;1;M1,q1,q2;(q1:1,3)(q2:2,3)(p1,M1:2,1)(p2,M1:1,4)(p3,M1:5,2)
  (p4,M1:4,3)(p5,M1:3,5)(p6:5,4)}
\end{lstlisting}
As a consequence of momentum conservation, $q_3 = - (q_1 + q_2)$ is not an independent external leg, and hence only two external legs are counted in the above example.

When performing a naive expansion in a scale with an external program like \texttt{exp}, the corresponding scale will not be listed in the list of scales. Furthermore an \texttt{,e} will be added to it in all relevant entries in the list of line momenta.
As an example consider the previous diagram, but in a Taylor expansion for a small internal quark mass \texttt{M1}:
\begin{lstlisting}[language=topsel]
{d2l3;6;2;2;1;q1,q2;(q1:1,3)(q2:2,3)(p1,M1,e:2,1)(p2,M1,e:1,4)
  (p3,M1,e:5,2)(p4,M1,e:4,3)(p5,M1,e:3,5)(p6:5,4)}
\end{lstlisting}
Similarly, a Taylor expansion for a large quark mass and small external momenta can be performed:
\begin{lstlisting}[language=topsel]
{d2l3;6;2;2;1;M1;(q1,e:1,3)(q2,e:2,3)(p1,M1:2,1)(p2,M1:1,4)(p3,M1:5,2)
  (p4,M1:4,3)(p5,M1:3,5)(p6:5,4)}
\end{lstlisting}

For programs such as \texttt{exp} a list of topology selection (\file{topsel}) entries can be generated. These take a similar form as the \file{edia} entries:
\begin{lstlisting}[language=topsel]
{Name; Number of lines; Number of loops; Number of independent ext. legs; 
  Number of masses; List of options; List of line momenta;
  Mass assignment}
\end{lstlisting}
The first five entries resemble those of an \file{edia} entry. They are followed by a list of options that can be filled for use with an external program. The last two entries contain
a list of line momenta and the mass assignment of each line. In contrast to the line momenta in \file{edia} entries the list of line momenta does not contain information regarding the masses.
This information is contained in a number with one digit per internal line. Each digit is either \texttt{0}, in case the line is massless, or a number between \texttt{1} and \texttt{9} corresponding to the masses \texttt{M1} to \texttt{M9} to be defined as \texttt{tapir.mass} option in the \file{config} file.

For example, a topology entry generated by \tapir resembling the \file{edia} entry of \texttt{d2l3} takes the form:
\begin{lstlisting}[language=topsel]
{Tri2la;6;2;2;1;;(q1:1,3)(q2:2,3)(p1:2,1)(p2:1,4)(p3:5,2)(p4:4,3)(p5:3,5)(p6:5,4);111110}
\end{lstlisting}

As in the case of the \file{dia} files, the format of the \file{edia} and \file{topsel} files is equivalent to those produced by \texttt{q2e} and used by \texttt{exp}.

\subsection{\file{topology} files}
\tapir can also generate \texttt{FORM} code to rewrite scalar products of loop momenta and propagators in terms of scalar functions.
Given a \file{topsel} entry, a subset of the line momenta is selected as the loop momenta and all other line momenta are decomposed as a sum
of external momenta and the loop momenta.

As an example, in the case of the \file{topsel} entry discussed above, the corresponding \texttt{FORM} code for rewriting numerators takes the form
\begin{lstlisting}[language=form]
* Reducible numerator momentum replacements
id p6 = -p3 + p5;
id p1 = p3 + q2;
id p2 = p3 + q1 + q2;
id p4 = p5 + q1 + q2;

.sort
\end{lstlisting}
Here \texttt{p3} and \texttt{p5} have been chosen as loop momenta.

In a next step, \tapir rewrites scalar products of loop momenta and external momenta in terms of scalar products of external momenta and inverse denominators.

For the case under consideration, the code is given by
\begin{lstlisting}[language=form]
* Numerator momentum product replacements
id p5.q1 = p4.p4/2 - p5.p5/2 - p5.q2 - q1.q1/2 - q1.q2 - q2.q2/2;
id p3.q1 = -p1.p1/2 + p2.p2/2 - q1.q1/2 - q1.q2;
id p3.q2 = p1.p1/2 - p3.p3/2 - q2.q2/2;
id p3.p5 = p3.p3/2 + p5.p5/2 - p6.p6/2;

.sort

* Define massive propagators
id p1.p1 = -1/s1m1 + M1^2;
id p2.p2 = -1/s2m1 + M1^2;
id p3.p3 = -1/s3m1 + M1^2;
id p4.p4 = -1/s4m1 + M1^2;
id p5.p5 = -1/s5m1 + M1^2;

.sort
\end{lstlisting}
Here, the first block of code rewrites scalar products in terms of the momenta of the propagators, while the second block takes into account masses of propagators.
The symbols \texttt{sImJ} denote the propagators $1/(m_j^2 - p_i^2)$.\footnote{This choice of the sign allows for direct use of the output through programs working with Euclidean loop momenta, such as \texttt{MATAD} or \texttt{MINCER}.}

In the next step, the massive denominator factors and remaining scalar products are rewritten in terms of a scalar function.

For the given example, the following \FORM code is generated:
\begin{lstlisting}[language=form]
* Combine to scalar topology function
id s5m1^n0? * s3m1^n1? * s4m1^n2? * s2m1^n3? * s1m1^n4? * 1/p6.p6^n5? * 1/p5.q2^n6? = 
   (-1)^n5 * (-1)^n6 * 
   Tri2la(n0,n1,n2,n3,n4,n5,n6);

.sort
\end{lstlisting}

Furthermore, if the problem involves a partial fraction decomposition, the corresponding code is appended here.

\subsection{\file{topology list} files}
\tapir also generates a \texttt{Mathematica} readable \file{topology list} file containing the definition of the propagators of
each resulting scalar topology function of the \file{topology} files. The entries have the form of
\begin{lstlisting}[language=C++]
{"Name", List of denominators, List of loop momenta}
\end{lstlisting}

In the case of the topology discussed above, the entry is given by
\begin{lstlisting}[language=C++]
{"Tri2la", {M1^2 - p5^2, M1^2 - p3^2, M1^2 - (p5 + q1 + q2)^2,
  M1^2 - (p3 + q1 + q2)^2, M1^2 - (p3 + q2)^2, -(p3 - p5)^2, -p5*q2},
  {p3, p5}}
\end{lstlisting}
Note that regular multiplication is used instead of a scalar product.

\section{\label{sec:usage}Usage and example}
The installation of \tapir is simple once the repository has been cloned. We recommend the usage of a pre-packed release version for a defined and well tested set of features.
New features can be found on the development (master) branch. The installation is done with

\begin{lstlisting}[language=shell]
$ make install
\end{lstlisting}
The shell-executable \file{tapir} can now be used. To get a hint on the command line options, the instruction

\begin{lstlisting}[language=shell]
$ /path/to/tapir --help
\end{lstlisting}

is useful. For further reference we strongly recommend the user manual in the \texttt{doc} folder of the repository.

To illustrate the basic features of \tapir we follow some basic examples which may occur in loop calculations. The first example highlights the multi-loop aspect and advantages of \tapir,
whereas the second and third examples show features which help to address more specific problems.
In the fourth example, we show how \tapir can be used to convert pre-generated Feynman rules into \tapir and \qgraf input files.
All examples can be found in the \texttt{example} subdirectory.

\subsection{Example 1: The 3-loop gluon self-energy}\label{subsec:threegluonex}

\subsubsection*{Generating symbolic diagrams}

Computations of this order are usually only possible with a suited \texttt{FORM} setup due to the large size of intermediate terms. Let us start with \texttt{qgraf} using the \tapir style file \file{qgraf-tapir.sty}.
Our model (called \file{lagrangian} file in \texttt{qgraf}) contains all QCD-relevant interactions with gluons, ghosts and quarks.
In addition, we split the four-gluon interaction into three sub-diagrams with only three-particle interactions using a so-called \textit{sigma particle} (see e.g.\ Ref.~\cite{Grozin:2005yg}).
This enables the factorization of color and Lorentz related parts of the Feynman rules, which we can hence compute separately for the whole diagram.

\begin{figure}
  \centering
  \includegraphics[width=1\textwidth]{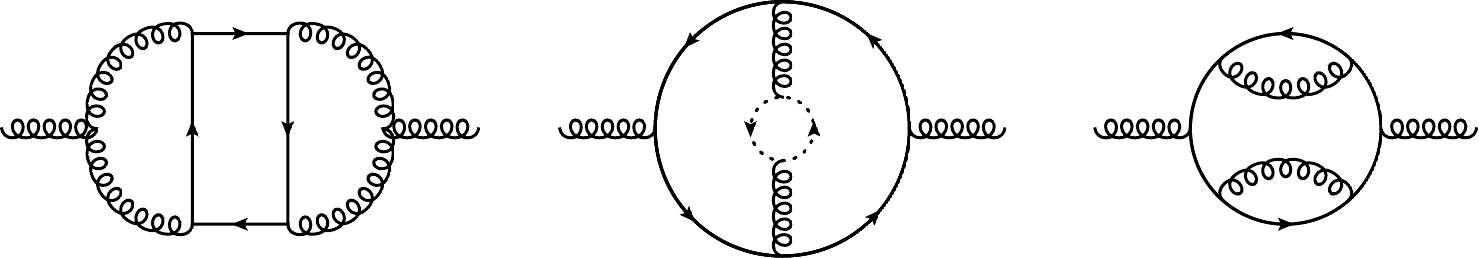}
  \caption{\texttt{qgraf} generates $\mathcal{O}(1600)$ diagrams for the gluon self-energy at 3-loop order. We included ghosts and sigma particles in our Lagrangian definition.}
  \label{fig:gg3l}
\end{figure}

\texttt{qgraf} produces $\mathcal{O}(1600)$ diagrams if all six quark flavors are treated individually. The resulting \file{qlist} file is a valid input for \tapir.
Example diagrams are shown in \autoref{fig:gg3l}.

To obtain a symbolic expression of each diagram, we have to specify Feynman rules for each propagator and interaction that was used in the \file{lagrangian} file.
The symbolic Feynman rules have to be provided in the \file{prop} and \file{vrtx} files, respectively. For example, the gluon propagator looks like

\begin{lstlisting}[language=vrtx]
{g,g:*Dg(<lorentz_index_vertex_1>,<lorentz_index_vertex_2>,<momentum>)
  |*prop(<colour_index_vertex_1>,<colour_index_vertex_2>)||}
\end{lstlisting}

A quark-gluon vertex is given by
\begin{lstlisting}[language=vrtx]
{fU,fu,g:*ffgVertex(<lorentz_index_particle_3>,
    <spinor_index_particle_1>,<spinor_index_particle_2>)
  |*GM(<colour_index_particle_3>,
    <spinor_index_particle_1>,<spinor_index_particle_2>)
  ||}
\end{lstlisting}

The structure of \file{prop} and \file{vrtx} entries is defined in \autoref{subsec:propvrtx}.

All configurations and options of \tapir can be provided in the \file{config} file. If we simply want to insert all Feynman rules per diagram, the \file{config} file reads:

\begin{lstlisting}[language=tapirconf]
Define propagator Feynman rules
* tapir.propagator_file qcd.prop

Define vertex Feynman rules
* tapir.vertex_file qcd.vrtx

Declare and assign masses to particles
* tapir.scales M1
* tapir.mass ft:M1

The following options can also be given via command line.
qgraf input
* tapir.qlist qlist.3

Output
* tapir.diaout gg3l.dia
* tapir.ediaout gg3l.edia
\end{lstlisting}

The options here are more or less self-explanatory. To specify massive particles the \texttt{tapir.mass} option for the corresponding particle needs to be provided. As an output we specify two files: \file{dia} and \file{edia} files are discussed in \Autoref{subsec:diafiles} and \Autoref{subsec:ediafiles}

The last three options can also be provided via the command line. But here we specify everything in the \file{config} file such that we can run \tapir simply with

\begin{lstlisting}[language=shell]
$ /path/to/tapir -c myconf.conf
\end{lstlisting}

\subsubsection*{Topology analysis}
\tapir is also able to map and minimize the topologies of the generated diagrams. For this, we add the following lines to the \file{config} file:

\begin{lstlisting}[language=tapirconf]
* tapir.minimize
* tapir.topselout reducedTopologies.topsel
\end{lstlisting}

\begin{figure}
  \centering
  \includegraphics[width=1\textwidth]{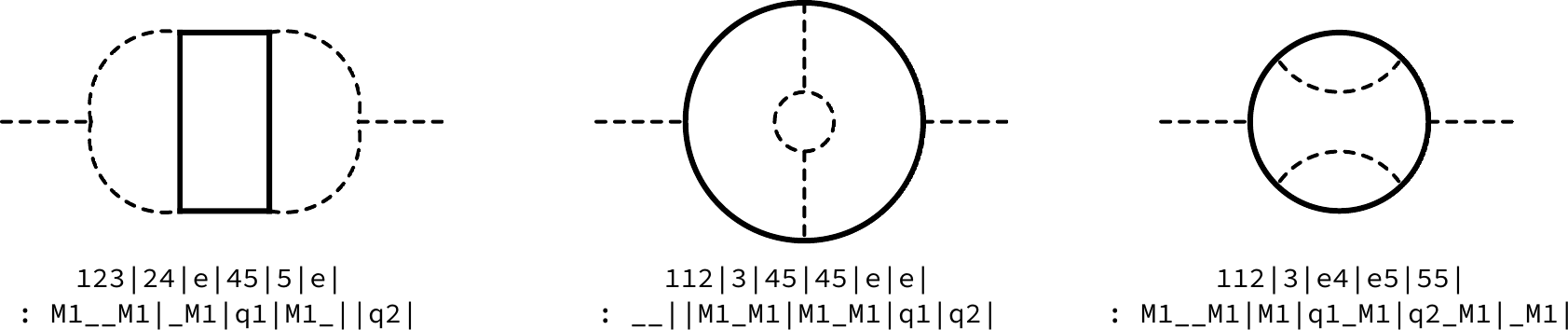}
  \caption{The topology structure of the diagrams of \autoref{fig:gg3l} shows that only the propagator connections and the masses are relevant for the Nickel index. The dashed lines indicate massless propagators and the solid lines massive ones.}
  \label{fig:gg3l-topos}
\end{figure}

The first option computes the Nickel index (see \autoref{sec:nickel}) for every diagram. Examples are shown in \autoref{fig:gg3l-topos}. Afterwards, the Nickel indices between all diagrams are compared and only unique ones are kept. The remaining topologies are then written to a \file{topsel} file which has a similar format as the \file{edia} file. 
The former is used to map onto target topologies with \texttt{exp}.
We adapted the file format here to express topologies in general. It is also possible to use a \file{topsel} file as input and, for example, minimize its content with respect to another \file{topsel} file. This explicit handling of topologies in in- and output allows for the realization of different minimization and mapping approaches that fit best to the problem at hand.

At the end of the minimization we are left with 60 unique topologies. The following option gives us the possibility to express diagrams in terms of a scalar topology function:

\begin{lstlisting}[language=tapirconf]
* tapir.topologyfolder topologies
\end{lstlisting}

It creates a \texttt{FORM} file for every remaining topology, which reduces the possible numerator and denominator structures assuming common Feynman rules (i.e.\ quadratic massive or massless denominators). More specialized propagators, e.g.\ eikonal ones, can also be requested by a different option. These \file{topology} files reside in the directory \texttt{topologies} together with a \texttt{Mathematica} readable \file{topology list} file which can be used as an input for reduction programs such as \texttt{FIRE}.
Alternatively, the topology information can also be written to a \texttt{YAML}-formatted file that can be used with integration-by-parts reduction programs such as \texttt{Reduze} \cite{Studerus:2009ye,vonManteuffel:2012np} or \texttt{Kira} \cite{Maierhofer:2017gsa,Klappert:2020nbg}.

For visualization the option 
\begin{lstlisting}[language=tapirconf]
* tapir.topologyart topologies.tex
\end{lstlisting}
can be used to generate a \texttt{TikZ-Feynman} representation for every remaining topology. The generated \LaTeX~file can be compiled using \texttt{lualatex}.

\subsection{Example 2: Partial fraction decomposition}
This example is concerned with the partial fraction decomposition of Feynman integrals with linearly dependent propagators. Such integrals occur whenever kinematic configurations are chosen in which
external momenta are nullified or become proportional to other external momenta. As an example consider QCD corrections to $B_s -\overline{B}_s$ mixing. This requires the computation of $\bar{b} + s \rightarrow b + \bar{s}$ in forward scattering diagrams (see e.g.\ Ref.~\cite{Beneke:1998sy}). If terms suppressed by the $b$-quark mass are neglected, these diagrams
can be computed for vanishing external strange-quark momentum. Thus, we can set $q_2 = q_4 = 0$ and $q_1 = -q_3$.
In the following we will use \tapir to identify all relevant topologies for this problem, nullify $q_2$ and $q_4$, find linearly independent topologies and generate \texttt{FORM} code to map
linearly dependent integrals onto them.

After generating the relevant four-point diagrams with \texttt{qgraf} we can instruct \tapir to nullify the two momenta using the options
\begin{lstlisting}[language=tapirconf]
* tapir.external_momentum q2:0
* tapir.external_momentum q4:0
\end{lstlisting}
in the configuration file (see \texttt{example/Bmix/Bmix.conf}).

Calling \tapir with
\begin{lstlisting}[language=shell]
$ /path/to/tapir -c Bmix.conf -q qlist.2 -m -t Bmix.topsel -f topologies  -pm
\end{lstlisting}
loads the configuration file \texttt{Bmix.conf} and the output of \texttt{qgraf} (\texttt{example/Bmix/qlist.2}). The \texttt{-m} switch leads to the minimization of integral topologies based on the given diagrams,
followed by the generation of a \file{topsel} file for \texttt{exp}. The \texttt{-f} switch instructs \tapir to generate \texttt{FORM} \file{topology} files for rewriting propagators and scalar products in terms of
scalar integral familes. Finally, the \texttt{-pm} switch triggers the partial fraction decomposition of linearly dependent denominator functions of the integral families and a subsequent minimization. The respective code is added to the \file{topology} files.

\subsection{Example 3: Filtering cuts}\label{subsec:cutfilters}
In computations based on the method of reverse unitarity \cite{Anastasiou:2002yz} phase space integrals over squared amplitudes are rewritten as cut loop integrals in forward scattering kinematics
(see \autoref{sec:cutkosky}).
This method allows the application of techniques developed for regular loop integrals to phase space integrals. However, diagram generators such as \texttt{qgraf} are not capable to only generate
diagrams with the correct cuts. As a consequence, we need to filter the output of \texttt{qgraf} using \tapir before doing further manipulations.

As an example consider higher order corrections to Higgs boson pair production in the large top quark mass expansion. In this expansion top quark loops reduce to effective gluon-Higgs vertices
and as a consequence only diagrams with massless quarks, gluons and Higgs bosons need to be considered. Example diagrams can be found in Refs.~\cite{Davies:2019xzc,Davies:2021kex}.
To filter the output of \texttt{qgraf} we add 
\begin{lstlisting}[language=tapirconf]
* tapir.filter cuts : true : q1,q2 : h : 2,2
* tapir.filter cuts : true : q1,q2 : g,c,fq : 1,2
\end{lstlisting}
to the respective configuration file. These options select all diagrams with an $s$-channel cut through two Higgs bosons (\texttt{h})
and any combination of one or two gluons (\texttt{g}), light quarks (\texttt{fq}) or ghosts (\texttt{c}).

Should we only be interest in final states with exactly two quarks we can modify the second line to read
\begin{lstlisting}[language=tapirconf]
* tapir.filter cuts : true : q1,q2 : fq : 2,2
\end{lstlisting}

Alternatively, we could filter diagrams with cuts through gluons and ghosts but no quarks by the following combination of options:
\begin{lstlisting}[language=tapirconf]
* tapir.filter cuts : true : q1,q2 : h : 2,2
* tapir.filter cuts : true : q1,q2 : g,c : 1,2
* tapir.filter cuts : false : q1,q2 : fq :
\end{lstlisting}

The option arguments are described as follows. The first boolean argument specifies whether the diagrams fulfilling the stated cutting restriction are exclusively kept (\texttt{true}) or excluded (\texttt{false}) from further evaluation.

The second argument describes which external momenta are treated as ``sources''. By specifying some external lines (according to their momenta) as sources, all other external lines are treated as sinks.

The third argument (if provided) specifies for which particles the cutting restriction shall apply. Thus, cuts through different particle types can be cumulatively accounted for. An empty argument is equivalent to specifying all occurring particles.

The last argument defines the ranges of how many cuts of the particles of the given kind are allowed. Several allowed ranges can be provided at the same time. Alternatively, the range $[1, \infty)$ can be specified by leaving the argument empty.

As shown, several filter options can be provided in a single \file{config} file to confine the filtered diagram subset even more. Hence, the filter system allows logical ``$\wedge$'' and ``$\neg$'' operations. A logical ``$\vee$'' is only partly supported but can be induced by multiple \file{config} files.

\subsection{Example 4: Using an UFO model to provide Feynman Rules}
Here, we give a short example of how to use the \UFOReader to import Feynman rules corresponding to the Standard Model \texttt{FeynRules} \UFO files that are provided together with \tapir. 

The \UFOReader requires a minimal configuration file (e.g.\ \texttt{example/UFO/SM/ufo.conf}), consisting of only two options
\begin{lstlisting}[language=tapirconf]
Directory containing the input UFO files
* tapir.ufo_indir example/UFO/SM/Standard_Model_UFO/

Directory for the output (.lag, .vrtx, .prop, .inc) files
* tapir.ufo_outdir example/UFO/SM/tapir_SM_UFO/
\end{lstlisting}
which specify the input and output directories, respectively.
\tapir is then called by simply executing
\begin{lstlisting}[language=shell]
path_to_tapir$ ./tapir ufo -c example/UFO/SM/ufo.conf
\end{lstlisting}
and will read the \UFO files and produce a number of output files for later use with \qgraf, \tapir and \FORM.\newline
Note that the \file{lagrangian} and \file{prop} files will have two particle components, a ``transversal'' and a ``longitudinal'' one, for each massive vector boson, cf.\,\ref{subsec:ufopropagators} for details.
Some vertices in the \file{vrtx} file carry an explicit gauge-parameter dependence in the \texttt{Lorentz} fold, and the four-gluon vertex involves the function \texttt{nonfactag()}, both of which are explained in \ref{subsec:ufovertices}.
In addition, in the files \file{UFOdecl.inc} and \file{UFOrepl.inc}, we find declarations and definitions of coupling constants in terms of parameters of the theory, which can be used in \FORM-based setups.\newline
An exemplary use case of the \UFOReader would be to first convert a \texttt{FeynRules} \texttt{UFO} module into the previously mentioned files, then use \qgraf to generate all diagrams for a given process with the \file{lagrangian} file as input. Then, use the Feynman rule definitions in the usual way using \tapir a second time.

\section{\label{sec:sum}Summary and Outlook}
We present \tapir, a program for processing multi-loop Feynman diagrams and working with Feynman integral families. \tapir allows
for identification and minimization of topologies, partial fraction decomposition of linearly dependent topologies, filtering diagrams allowing for
any given Cutkosky cut, and it provides an interface to \texttt{UFO} for importing Feynman rules. Furthermore, \FORM code can be generated for the symbolic computation of
Feynman diagrams. We provide four examples showcasing the features of \tapir and discuss the relevant input and output files.
\tapir already came to use in~\cite{Gerlach:2021xtb}.

The source code, further examples as well as more detailed documentation can be obtained from \url{https://gitlab.com/tapir-devs/tapir}.
Furthermore, additional features and performance improvements might be added to \tapir in the future.

\section*{Acknowledgements}
We thank Fabian Lange, Vladyslav Shtabovenko and Matthias Steinhauser for careful reading of the manuscript.

This research was supported by the Deutsche Forschungsgemeinschaft (DFG, German Research Foundation)
under grant 396021762 — TRR 257 ``Particle Physics Phenomenology after the Higgs
Discovery''.
M.L.\ is supported by the BMBF grant 05H15VKCCA.

F.H.\ acknowledges the support of the Alexander von Humboldt Foundation.
This document was prepared using the resources of the Fermi National Accelerator Laboratory (Fermilab), a U.S. Department of Energy, Office of Science, HEP User Facility.
Fermilab is managed by Fermi Research Alliance, LLC (FRA), acting under Contract No. DE-AC02-07CH11359.

\appendix
\section{The \UFOReader module}\label{sec:UFOReader}
Given a representation of the Feynman rules of a model in terms of \file{vrtx}, \file{prop}, and \file{lagrangian} files, \tapir is able to generate \FORM expressions from \qgraf-generated~\cite{Nogueira:1991ex} diagrams.
The \file{vrtx}, \file{prop}, and \file{lagrangian} files can be manually written by the user.
If the model at hand is complicated, however, the number and complexity of Feynman rules grows, and such an approach can be prone to errors.
\texttt{FeynRules} \cite{Alloul:2013bka} is a convenient tool to automatically generate Feynman rules for a given theory.
The \UFOlong \cite{Degrande:2011ua} (\UFO) provides a complete, easy-to-use set of \texttt{FeynRules}-generated Feynman rules in terms of \python files.
In this section, we describe the \UFOReader module that allows to read in \UFO files and convert them into appropriate \file{vrtx}, \file{prop}, and \file{lagrangian} files.

The \python module produced by \UFO contains a number of different files with information about the particles (\file{particles.py}) and vertices (\file{vertices.py}, \file{couplings.py} and \file{lorentz.py}) of the given model, with additional information about the parameters given in \file{parameters.py}.
An \UFO module also contains the file \file{propagators.py} with information about spinless, spin-$\frac{1}{2}$ and spin-$1$ propagators, either in Feynman ($\xi = 1$) or unitary ($\xi \to \infty$) gauge.
We emphasize here that the \UFOReader should be used with models exported in the Feynman gauge $\xi = 1$.
Note that currently \UFO files are produced as \python~$2$ modules; in order to use such a module, the user has to convert it into a \python~$3$ module.
At present, the program \texttt{2to3}~\cite{python-2to3} seems to work well.

The \UFOReader module will import the model package from the path that is specified in the config file (e.g.\ \file{example/UFO/SM/ufo.conf}) with the directive
\begin{lstlisting}
* tapir.ufo_indir PATH_TO_MODEL
\end{lstlisting}
where \texttt{PATH\_TO\_MODEL} is the path to the directory containing the \file{\_\_init\_\_.py} file of the model.
In addition, a directory \texttt{PATH\_TO\_OUTPUT\_DIR} has to be given for the resulting files to be placed in, which can be specified with
\begin{lstlisting}
* tapir.ufo_outdir PATH_TO_OUTPUT_DIR
\end{lstlisting}
\tapir will then import and parse the \UFO files and produce the \file{lagrangian}, \file{prop} and \file{vrtx} files that can be used with \qgraf and \tapir again, as well as two additional files \file{UFOdecl.inc} and \file{UFOrepl.inc}, intended for the use with \FORM-based setups to take care of automated declaration and replacement of symbols, respectively.

In order to treat massive gauge bosons in general $R_\xi$ gauge, a list of massive spin-$1$ bosons is extracted from the list of all particles.
We will describe the treatment of massive vector bosons in \ref{subsec:ufopropagators}.
Note that the \UFOReader and resulting \file{prop} and \file{vrtx} files inherit the use of spinor, Lorentz and color indices from the original \UFO files.

The \UFOReader module has been tested on models containing spin-$0$, spin-$\frac{1}{2}$ as well as spin-$1$ particles in the trivial, fundamental or adjoint representation of $\mathrm{SU}(3)_\mathrm{C}$, respectively, and has been used in the context of calculations within the Standard Model and the Two-Higgs-Doublet Model.
Higher spin particles and particles in different representations of $\mathrm{SU}(3)_\mathrm{C}$ have also been implemented, but the primary use case is for the Standard Model and sufficiently similar theories.

The functions appearing in the Feynman rules are defined in the user documentation.
In the following we focus on two special cases that need further discussion: propagators of massive vector bosons and vertices for which the color and Lorentz structure does not factorize.

\subsection{Propagators}\label{subsec:ufopropagators}
The propagators of massless particles, massive spin-zero particles and spin-$\frac{1}{2}$ fermions as well as ghosts are translated into corresponding expressions suitable for \file{prop} files in a straightforward manner.
Massive gauge-boson propagators in $R_\xi$ gauge are cumbersome to deal with, since the propagator contains two denominators with different masses, $k^2 - M^2$ and $k^2 - \xi M^2$, at the same time, which makes the automated calculation of loop integrals rather cumbersome.
A standard solution is two split the massive propagator into two separate propagating particles, the ``transversal'' and ``longitudinal'' components, which is implemented in the \UFOReader module.
The $\xi$-dependent propagators of massive gauge bosons are expanded as
\begin{equation}
D_{\mu\nu} (k,M,\xi) = \mathcal{D}_{\mathrm{T},\mu\nu} (k, M, \xi) + \mathcal{D}_{\mathrm{L},\mu\nu} (k, M, \xi) \,,
\end{equation}
where the two components
\begin{align}
\mathcal{D}_{\mathrm{T},\mu\nu} (k, M, \xi) &= \frac{- \mathrm{i} \left(g_{\mu\nu} - \frac{k_\mu k_\nu}{M^2} \right)}{k^2 - M^2}~, \label{eq:massivegaugetran}\\
\mathcal{D}_{\mathrm{L},\mu\nu} (k, M, \xi) &= -\frac{\mathrm{i}}{M^2} \frac{k_\mu k_\nu}{k^2 - \xi M^2} \, \label{eq:massivegaugelong}
\end{align}
each feature only a single mass in the propagator denominator.
The subscripts $\mathrm{T}$ (``transversal'') and $\mathrm{L}$ (``longitudinal'') refer to the fact that 
\begin{equation}
k^\mu \mathcal{D}_{\mathrm{T},\mu \nu} \left(k, M, \xi \right) = 0
\end{equation}
if $k^2 = M^2$, that is if the massive gauge boson goes on-shell.
At the technical level, this involves introducing two particles $X_\mathrm{T}$ and $X_\mathrm{L}$ for each massive vector boson $X$ in the theory.
The corresponding propagators \texttt{Dtran} and \texttt{Dlong} have to be replaced by the user with their setup-dependent implementation of \Autoref{eq:massivegaugetran} and \Autoref{eq:massivegaugelong} in a later step.
In particular, it is easy to return to a specific choice of gauge, e.g.\ by simply setting \texttt{Dlong} to zero and 
\begin{equation}
\mathcal{D}_{\mathrm{T},\mu\nu} (k, M, \xi=1) = \frac{-\mathrm{i} g_{\mu\nu}}{k^2 - M^2}
\end{equation}
in the case of Feynman gauge $\xi = 1$.

Apart from this conceptually important aspect, the \UFOReader acts essentially as a \python-to-\tapir parser, i.e.\ it makes some notational changes to the objects and fields that appear.

\subsection{Vertices}\label{subsec:ufovertices}
Similarly to the propagators, the vertices also almost exclusively undergo some parsing, with two noteworthy exceptions.

The first point concerns the general $R_\xi$ gauge that is restored by the \UFOReader.
In order to restore the correct gauge dependence of vertices, a few vertices have to carry an overall factor of the corresponding gauge parameter $\xi$~\cite{Romao:2012pq}. 
This factor is simply prefixed to the \texttt{Lorentz} fold string of the corresponding vertex in the \file{vrtx} file.
As an example, consider the vertex of the ghost corresponding to the $Z$-boson, the ghost corresponding to the $W^+$-boson and the negatively charged goldstone boson.
This vertex is proportional to the gauge parameter of the $Z$-boson and takes the form
\begin{lstlisting}[language=vrtx]
{cghZ,CghWp,Gp: *<gauge_parameter_xiZ> *( ufoGC95 * ( 1 )) | *(1) | | }
\end{lstlisting}
In addition, since massive vector bosons are split into two parts (see \ref{subsec:ufopropagators}), for each possible combination of ``transversal'' and ``longitudinal'' propagator components entering a vertex, an identical copy of the vertex Feynman rule is generated.

Secondly, the factorisation of a Feynman rule into a \texttt{Lorentz} part and a \texttt{QCD} part is not always trivial, e.g.\ the four-gluon vertex of QCD.
A common approach is to introduce an auxiliary \textit{sigma particle} (cf. \autoref{subsec:threegluonex}) and write the four-gluon vertex as a linear combination of several sigma-exchange diagrams.
In the \UFOReader, we follow a less QCD-tailored approach, based on Ref.~\cite{Davies:2019esq}. In Feynman rules of non-factorising vertices we introduce an auxiliary function \texttt{nonfactag()} with three arguments.
For example, the rule for the four-gluon vertex reads
\begin{lstlisting}[language=vrtx]
{g,g,g,g:*( ufoGC12 * ( 
    d_(<lorentz_index_particle_1>,<lorentz_index_particle_4>)
    *d_(<lorentz_index_particle_2>,<lorentz_index_particle_3>)
    -d_(<lorentz_index_particle_1>,<lorentz_index_particle_2>)
    *d_(<lorentz_index_particle_3>,<lorentz_index_particle_4>) 
  )*nonfactag(0,1,<local_index_F>)
+  ufoGC12 * ( 
    d_(<lorentz_index_particle_1>,<lorentz_index_particle_4>)
    *d_(<lorentz_index_particle_2>,<lorentz_index_particle_3>) 
    -d_(<lorentz_index_particle_1>,<lorentz_index_particle_3>)
    *d_(<lorentz_index_particle_2>,<lorentz_index_particle_4>) 
  ) *nonfactag(0,0,<local_index_F>)
+  ufoGC12 * ( 
    d_(<lorentz_index_particle_1>,<lorentz_index_particle_3>)
    *d_(<lorentz_index_particle_2>,<lorentz_index_particle_4>) 
    -d_(<lorentz_index_particle_1>,<lorentz_index_particle_2>)
    *d_(<lorentz_index_particle_3>,<lorentz_index_particle_4>) 
  ) *nonfactag(0,2,<local_index_F>))
|*( ufocomplex(0,1)
  *V3g(<colour_index_particle_-1>,<colour_index_particle_1>,
    <colour_index_particle_2>)*ufocomplex(0,1)
  *V3g(<colour_index_particle_3>,<colour_index_particle_4>,
    <colour_index_particle_-1>)*nonfactag(0,0,<local_index_F>)
+ ufocomplex(0,1)
  *V3g(<colour_index_particle_-1>,<colour_index_particle_1>,
    <colour_index_particle_3>)*ufocomplex(0,1)
  *V3g(<colour_index_particle_2>,<colour_index_particle_4>,
    <colour_index_particle_-1>)*nonfactag(0,1,<local_index_F>)
+ ufocomplex(0,1)
  *V3g(<colour_index_particle_-1>,<colour_index_particle_1>,
    <colour_index_particle_4>)*ufocomplex(0,1)
  *V3g(<colour_index_particle_2>,<colour_index_particle_3>,
    <colour_index_particle_-1>)*nonfactag(0,2,<local_index_F>))
||}
\end{lstlisting}
There are three terms inside the \texttt{Lorentz} fold, each coming with a factor of \texttt{nonfactag()}.
These factors differ in the second argument of \texttt{nonfactag()}, which labels the term inside the Feynman rule, but share
the same third argument that is replaced by an index, unique to the position of the vertex in a given Feynman diagram by \tapir. The \texttt{QCD} fold is structured in the same way.
The first argument of \texttt{nonfactag()} counts the number of different non-factorising Feynman rules, starting from zero.
When computing diagrams involving a non-factorising vertex, both, the color factor and the \texttt{Lorentz} part, can be calculated independently. When multiplying both results, we can
set products of \texttt{nonfactag()} to $0$ if they correspond to the same vertex but different terms or to $1$ if they correspond to the same vertex and the same term. This can be achieved by two simple \texttt{FORM} replacement rules:
\begin{lstlisting}[language=form]
*** First set matching products to 1
id nonfactag(a?,b?,c?)*nonfactag(a?,b?,c?) = 1;
*** All remaining products corresponding to the same vertex vanish
id nonfactag(a?,b?,c?)*nonfactag(d?,e?,c?) = 0;
\end{lstlisting}

\section{\texttt{YAML} support}\label{sec:yaml}
Certain input and output files can also be given to \tapir in \texttt{YAML} format.
As a minimal example, we will show how the configuration file can be passed as \texttt{YAML} files.
For a complete documentation, we refer the reader to the online resources.

The \texttt{YAML} configuration option names are the same as in the configuration file style presented above, with the exception that all prefixes \texttt{* tapir.} are absent.
Options that are associated with a single value are written in the \texttt{key:} \texttt{value} format, with simple flags such as \texttt{* tapir.euclidean} receiving an extra boolean value \texttt{True}.
For options with comma-separated values, the notation \texttt{option: [value1, value2, ...]} is used, while options with colon-separated values are passed e.g.\ using
\begin{lstlisting}[language=yaml]
mass:
  - {ft: M1, Wp: M2, Z: M3}
\end{lstlisting}
Returning to the configuration file presented in \autoref{subsec:threegluonex}, its \texttt{YAML} equivalent (without comments) takes the following form:
\begin{lstlisting}[language=yaml]
config:
  propagator_file: qcd.prop
  vertex_file: qcd.vrtx
  qlist: qlist.3
  scales: [M1]
  mass:
    - ft: M1
  minimize: True
  diaout: gg3l.dia
  ediaout: gg3l.edia
  topselout: reducedTopologies.topsel
  topologyfolder: topologies
  topologyart: topologies.tex
\end{lstlisting}
The \texttt{tapir.filter} option has the most complicated structure, which is also mirrored in the way it must be implemented in a \texttt{YAML} configuration file.
For example, the last filter configuration discussed in \autoref{subsec:cutfilters} takes the form
\begin{lstlisting}[language=yaml]
config:
  # other config options
  # ...
  filter:
    - type: cuts
      value: True
      sources: [q1,q2]
      particles: h
      intervals: [2,2]
    - type: cuts
      value: True
      sources: [q1,q2]
      particles: [g,c]
      intervals: [1,2]
    - type: cuts
      value: False
      sources: [q1,q2]
      particles: fq
\end{lstlisting}

Apart from the configuration file, \tapir is also able to read and write \texttt{YAML}-formatted diagram (\file{dia}) and \file{topology} files.
The file \file{qgraf-yaml.sty} contained in the \tapir repository can be used with \qgraf should the user wish to obtain \texttt{YAML}-style \qgraf output.

\bibliographystyle{elsarticle-num}
\bibliography{literature}
\end{document}